\documentclass[3p,times,procedia]{elsarticle}
\usepackage{nupha_ecrc}

\volume{00}

\firstpage{1}

\journalname{Nuclear Physics A}

\runauth{}

\jid{nupha}

\jnltitlelogo{Nuclear Physics A}

\usepackage{amssymb}
\usepackage{amsthm}
\usepackage{amsmath}
\usepackage{hyperref}
\hypersetup{colorlinks=true}
\usepackage{graphicx}
\graphicspath{{fig/}}
\usepackage{subcaption}
\newcommand{\trento}{T\raisebox{-0.5ex}{R}ENTo}

\usepackage[figuresright]{rotating}

\newcommand{\be}{\begin{eqnarray}}
\newcommand{\ee}{\end{eqnarray}}

\newcommand{\Tr}{\mathrm{Tr}}

\begin{document}

\begin{frontmatter}



\dochead{XXVIIIth International Conference on Ultrarelativistic Nucleus-Nucleus Collisions\\ (Quark Matter 2019)}

\title{Quarkonium Production in Heavy Ion Collisions: From Open Quantum System to Transport Equation}


\author[label1,label2]{Xiaojun Yao}
\ead{xjyao@mit.edu}
\author[label3,label2]{Weiyao Ke}
\author[label2]{Yingru Xu}
\author[label2]{Steffen A. Bass}
\author[label2]{Thomas Mehen}
\author[label2]{Berndt M\"uller}
\address[label1]{Center for Theoretical Physics, Massachusetts Institute of Technology, Cambridge, MA, 02139, USA}
\address[label2]{Department of Physics, Duke University, Durham, NC 27708, USA}
\address[label3]{Nuclear Science Division, Lawrence Berkeley National Laboratory, Berkeley, CA 94720, USA}

\begin{abstract}
Using the open quantum system formalism and effective field theory of QCD, we derive the Boltzmann transport equation of quarkonium inside the quark-gluon plasma. Our derivation illuminates that the success of transport equations in quarkonium phenomenology is closely related to the separation of scales in the problem.
\end{abstract}

\begin{keyword}
quarkonia, open quantum system, Boltzmann transport equation, quark-gluon plasma, heavy ion collisions


\end{keyword}

\end{frontmatter}


\section{Introduction}
\label{sect:intro}
Heavy quarkonium has been used an important probe of the quark-gluon plasma (QGP) in heavy ion collisions since the early study of the plasma static screening effect on the heavy quark bound state \cite{Matsui:1986dk}. At sufficiently high temperature, the attractive potential between the heavy quark pair ($Q\bar{Q}$) is significantly suppressed and becomes too weak to support the formation of the bound state.
The melting temperature is ordered by the size of the state, with the smallest state having the highest melting temperature. In other words, one expects a sequential melting.

However, this simple picture is complicated by other medium effects. Another plasma effect, dynamical screening, describes the dissociation of quarkonium in a collision process. This process generates a thermal width of the state that increases with temperature. The inverse process of dissociation, recombination, has been proposed long time ago \cite{Thews:2000rj,Andronic:2007bi} and shown to be a crucial production mechanism of charmonium in heavy ion collisions. Therefore, phenomenological studies need to account for them consistently.

\begin{figure}[h]
    \centering
     \includegraphics[height=0.5in]{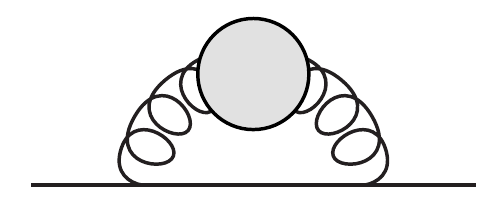}
     \caption{Schematic diagram of the thermal loop correction to the quarkonium propagator.}
     \label{fig:loop}
\end{figure}

Transport equations with both screening and recombination have been applied successfully in quarkonium phenomenology in heavy ion collisions. Both plasma screening effects can be studied from the thermal loop correction to the quarkonium propagator. A schematic diagram is shown in Fig.~\ref{fig:loop}. The real and imaginary parts of the $Q\bar{Q}$ potential can be extracted from the propagator. The real part encodes the static screening while the imaginary part is related to the dissociation rate. However, currently, most recombination calculations depend on various models: statistical recombination model, coalescence model and a model based on detailed balance and thus lack similar theoretical understanding as the screening effect.

In this proceeding, using the open quantum system formalism, we will demonstrate a consistent theoretical framework for both screening and recombination. Furthermore, using the potential nonrelativistic QCD (pNRQCD) \cite{Brambilla:1999xf,Fleming:2005pd}, which has been applied to study quarkonium dissociation and transport as an open system \cite{Brambilla:2011sg,Brambilla:2016wgg}, we will derive the Boltzmann transport equation from QCD. Finally, we will show some phenomenological results calculated from the derived equation. The proceeding is organized as follows: in Sect.~\ref{sect:oqs}, we will explain the open quantum system formalism and the derivation of the transport equation. Phenomenological results will be shown in Sect.~\ref{sect:pheno}. Conclusions will be drawn in Sect.~\ref{sect:conclu}.

\section{Open Quantum System and Transport Equation}
\label{sect:oqs}
We consider a total system consisting of a sub-system and an environment. More specifically, the sub-system consists of $Q\bar{Q}$ pairs that can be bound or unbound while the environment is the QGP in local thermal equilibrium. The total system is closed and evolves unitarily. When we focus on the dynamics of the sub-system, the environment degrees of freedom are traced out. So the sub-system is an open system and evolves non-unitarily. The evolution of the open system is also time-irreversible even though the underlying theory of the total system is time-reversible. The evolution of the sub-system density matrix $\rho_S$ can be written as
\be
\label{eqn:1}
\rho_S(t) = \Tr_E \{ U(t,0) \rho(0) U^\dagger(t,0) \}\,,
\ee
where $\rho(0)$ is the total density at $t=0$ and $U(t,0)$ is the time evolution operator of the total system.

Two limits exist where the evolution equation of the sub-system can be simplified: quantum optical limit and quantum Brownian motion. Whether to take one limit or the other depends on the time scales in the problem. Three time scales are relevant here: the system intrinsic time scale $\tau_S$, the environment correlation time $\tau_E$ and the system relaxation time $\tau_R$. The quantum optical limit corresponds to the hierarchy $\tau_R \gg \tau_E$ and $\tau_R \gg \tau_S$ while the limit of quantum Brownian motion is valid when $\tau_R \gg \tau_E$ and $\tau_S \gg \tau_E$. For quarkonia transport inside QGP, $\tau_S$ is the typical time scale that the pair is revolving around each other, $\tau_R$ is the time scale of approaching detailed balance at equilibrium. Both limits require $\tau_R \gg \tau_E$, which means the environment correlation has lost between two changes of the sub-system. In other words, the sub-system evolution is Markovian (no memory effect). For the quantum optical limit, $\tau_R \gg \tau_S$ indicates that during the $Q\bar{Q}$-medium reaction time, the pair has revolved around each other for many periods. Only then does it make sense to study bound state formation and dissociation. If the pair has not finished one period of revolving before the dissociation, one cannot treat the pair as a well-defined bound state. It is more like two independent heavy quarks. For the quantum Brownian motion,  $\tau_S \gg \tau_E$ guarantees that the sub-system is only sensitive to the low-frequency part of the environment correlation.

We will work in the quantum optical limit and justify the hierarchy of time scales now. We will use a version of pNRQCD that is valid when $M\gg Mv \gg Mv^2 \gtrsim T$, where $M$ is the heavy quark mass, $v$ is the typical relative velocity between the pair and $T$ is the temperature of the QGP, determining the environment correlation time $\tau_E \sim \frac{1}{T}$. Here $\frac{1}{Mv}$ is the typical size of quarkonium while $\frac{1}{Mv^2}$ is the intrinsic time scale of the pair inside quarkonium. For both charmonium and bottomonium, $Mv^2\sim 500$ MeV, which is roughly on the order of the highest temperature achieved in current heavy ion collision experiments. If one assumes the scale $Mv$ is perturbative (which works better for bottomonium), one can construct pNRQCD by perturbative matching calculations. But the arguments given below also work for the case of non-perturbative matching. In pNRQCD, quarkonium is described as a bound color singlet. Unbound $Q\bar{Q}$ pairs can be color singlet or cotet. At the linear order in the multipole expansion, the singlet and octet interact via a dipole interaction with a chromo-electric field. So for the quarkonium dissociation and recombination, the relevant scattering vertex scales as $rT$ where $r$ is the typical size of the quarkonium. Thus the interaction between the sub-system and the environment is weak: The vertex scales as $\frac{T}{Mv} \lesssim v$, which is assumed to be small for nonrelativistic heavy quarks. The sub-system relaxation time can be estimated as $(rT)^2T \lesssim v^2T \ll T \lesssim Mv^2$. Taking the inverse, we justify the Markovian approximation $\tau_R \gg \tau_E$ and also $\tau_R \gg \tau_S$.

\begin{figure}[h]
    \centering
     \includegraphics[height=0.8in]{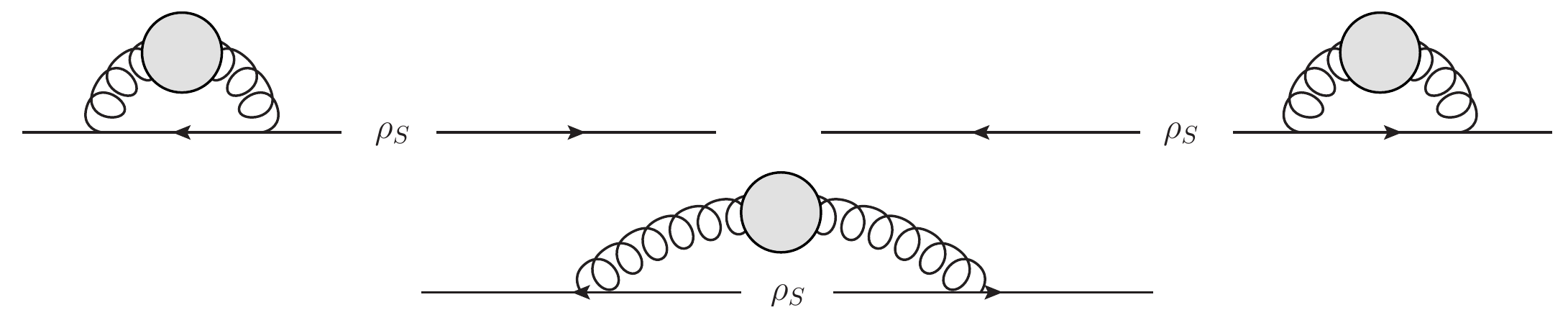}
     \caption{Schematic diagram of the thermal loop correction of the time evolution of the sub-system density matrix. The arrows indicate the time evolving direction.}
     \label{fig:lindblad}
\end{figure}

In the interaction picture, assuming $\rho(0) = \rho_S(0) \otimes \rho_E$ and expanding the evolution operator $U(t,0)$ in (\ref{eqn:1}) to second order in perturbation leads to the Lindblad equation
\be
\label{eqn:lindblad}
\rho_S(t) = \rho_S(0)  -i\sum_{a,b} \sigma_{ab}(t) [L_{ab}, \rho_S(0)] + \sum_{a,b,c,d} \gamma_{ab,cd} (t) \Big( L_{ab}\rho_S(0)L^{\dagger}_{cd} - \frac{1}{2}\{ L^{\dagger}_{cd}L_{ab}, \rho_S(0)\}  \Big) \,,
\ee
where detailed definitions of each term can be found in Ref.~\cite{Yao:2018nmy}.
Here we will explain each term diagrammatically, as shown in Fig.~\ref{fig:lindblad}. The $\sigma_{ab}(t) [L_{ab}, \rho_S(0)]$ term and the $- \frac{1}{2} \gamma_{ab,cd} \{ L^{\dagger}_{cd}L_{ab}, \rho_S(0)\} $ term come from the first two diagrams, which are just the propagators of quarkonium if the sub-system density matrix is a quarkonium state. They correspond to the real and imaginary parts and thus represent the static screening and dissociation, as discussed above in Sect.~\ref{sect:intro}. The $ \gamma_{ab,cd}L_{ab}\rho_S(0)L^{\dagger}_{cd}$ term comes from the third diagram. It is this diagram with the cross-talk between the two evolution operators that represents recombination, which becomes manifest only if we study the density matrix evolution.

Finally, if we take the Wigner transform of the sub-system density matrix, we can reproduce the Boltzmann transport equation \cite{Yao:2018nmy,Yao:2019jir}. This explains why the transport equation approach is successful in quarkonium phenomenology in heavy ion collisions: The equation can be systematically derived from QCD under the separation of scales: $M\gg Mv \gg Mv^2 \gtrsim T$, which is approximately valid in current heavy ion collision experiments. 

\section{Phenomenological Results}
\label{sect:pheno}
In this section, we will show phenomenological results based on the quarkonium transport equation derived in the last section. After quarkonium dissociation, a valid description of the in-medium heavy quark dynamics is the transport equation of open heavy quarks \cite{Ke:2018tsh}. Thus, we have to couple the transport equations of quarkonium with those of open heavy quarks \cite{Yao:2018zrg,Yao:2018sgn}. Detailed balance and kinetic thermalization of quarkonium in a QGP box with a constant temperature have been demonstrated by using the coupled transport equations \cite{Yao:2017fuc}. 

\begin{figure}[h]
    \centering
        \begin{subfigure}[t]{0.33\textwidth}
        \centering
        \includegraphics[height=1.4in]{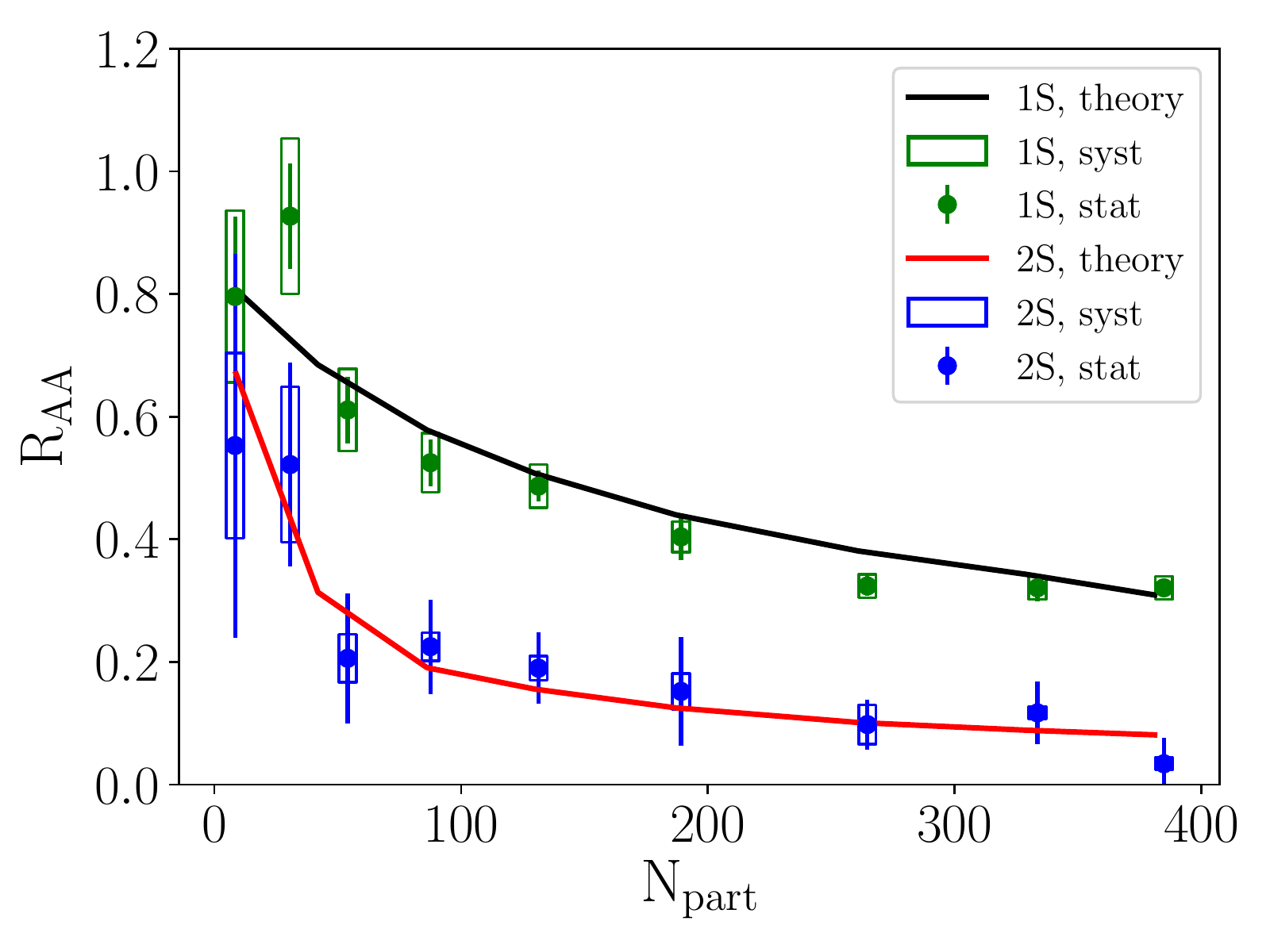}
        \caption{$R_{AA}$ as a function of centrality.}
    \end{subfigure}%
    ~
    \begin{subfigure}[t]{0.33\textwidth}
        \centering
        \includegraphics[height=1.4in]{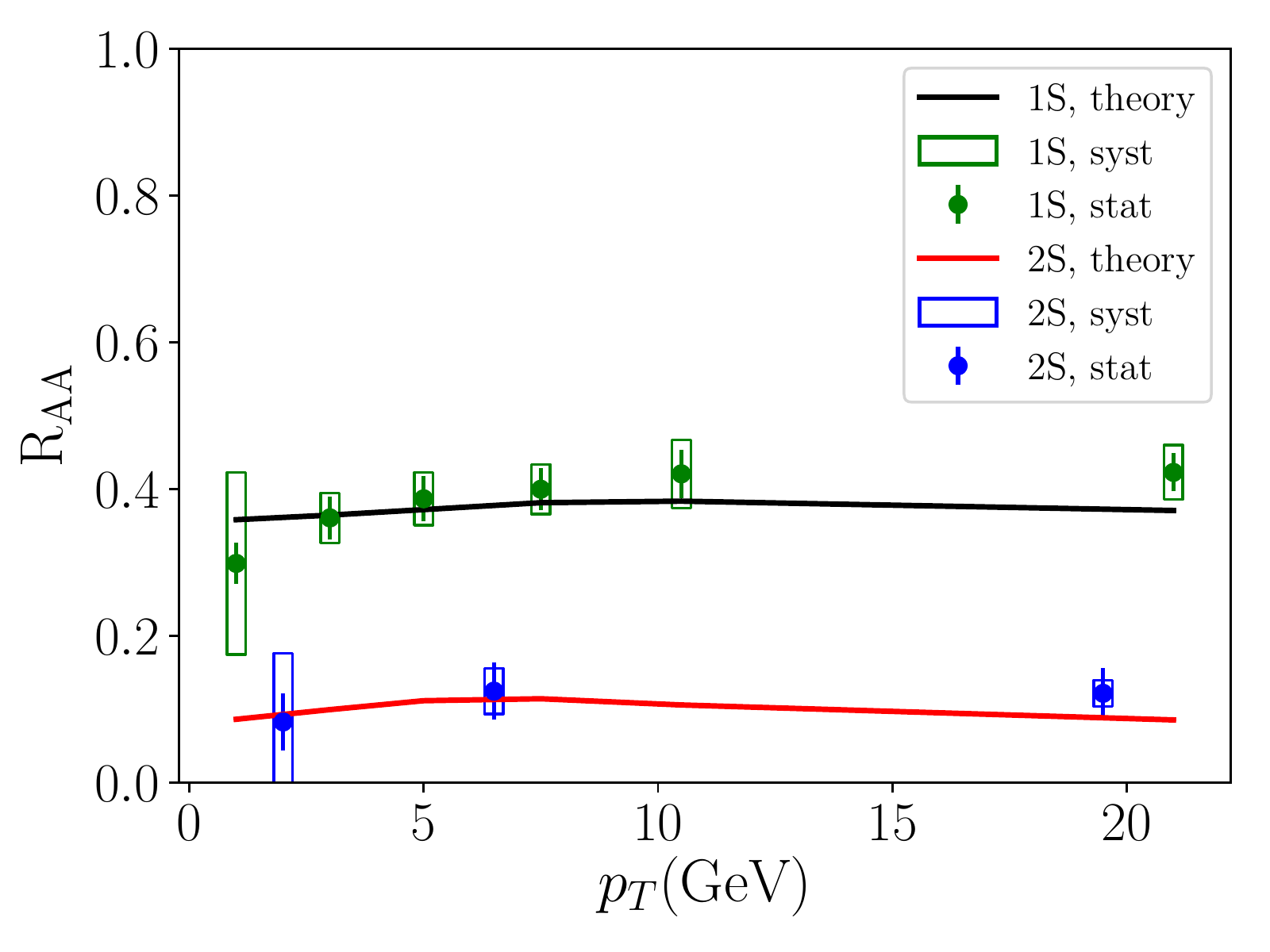}
        \caption{$R_{AA}$ v.s. $p_T$ in $0\%-100\%$ centrality.}
    \end{subfigure}%
    ~
    \begin{subfigure}[t]{0.33\textwidth}
        \centering
        \includegraphics[height=1.4in]{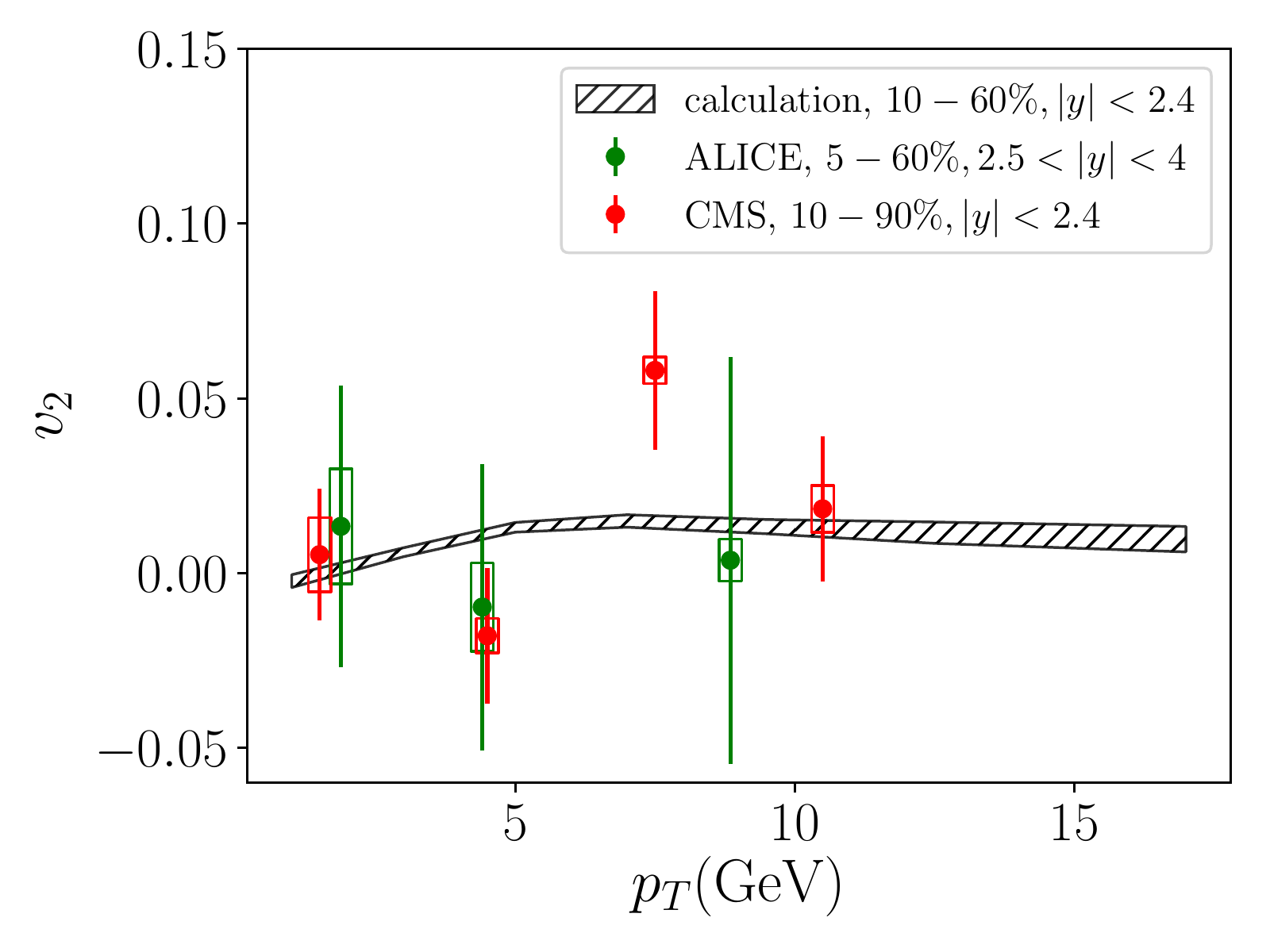}
        \caption{$v_2$ of $\Upsilon$(1S).}
    \end{subfigure}%
    \caption{$R_{AA}$ and $v_2$ in $5.02$ TeV Pb-Pb collisions. Data are taken from Refs.~\cite{Sirunyan:2018nsz,CMS:2019uhg,Acharya:2019hlv}.}
    \label{fig:cms}
\end{figure}

We will focus on the $\Upsilon$ production and include $\Upsilon$(1S) and $\Upsilon$(2S) in the coupled transport equations. We will only show results for $5.02$ TeV Pb-Pb collisions here. Results for other collision systems and energies have been reported in Ref.~\cite{Yao:2018zrg}.
The melting temperature of $\Upsilon$(2S) is set to be $210$ MeV. The coupling constant is fixed to be $\alpha_s=0.3$ at the scale $Mv$ and the singlet attractive potential is $V_S=-\frac{0.56}{r}$. The initial momenta of particles produced from hard scattering are sampled from \textsc{Pythia} \cite{Sjostrand:2014zea} with nuclear parton distribution function \cite{Eskola:2009uj}. The positions of initial hard scattering vertices are sampled from the binary collision density profile calculated in \trento~\cite{Moreland:2014oya}. {\trento} also generates the initial entropy density for the 2+1D relativistic viscous hydrodynamic evolution \cite{Shen:2014vra}. The initial condition and medium properties have been calibrated to the soft hadron observables \cite{Bernhard:2016tnd}. We will use an event-averaged hydrodynamic medium. The branching ratio of $\Upsilon$(2S) to $\Upsilon$(1S) in hadronic phase is $0.26$. The comparison of $R_{AA}$ and $v_2$ between the calculation and the measurements are shown in Fig.~\ref{fig:cms}. Our calculations can describe the data, though the current uncertainty in the $v_2$ measurements is still large.

\section{Conclusions}
\label{sect:conclu}
In this proceeding, we argued that the success of transport equations in quarkonium phenomenology in heavy ion collisions is related to the separation of scales and explained the derivation of transport equation in QCD. We also showed some phenomenological results that are consistent with experiment measurements. In the future, we will include other excited bottomonium states such as 1P, 2P and 3S states. Generalization to the study of doubly charmed baryons can be found in Ref.~\cite{Yao:2018zze}.

\section*{Acknowledgments}
This work is supported by U.S. Department of Energy research grants DE-FG02-05ER41367 and DE-FG02-05ER41368. XY also acknowledges support from the grant DE-SC$0011090$, Brookhaven National Laboratory and Department of Physics, Massachusetts Institute of Technology.





\bibliographystyle{elsarticle-num}



\end{document}